\documentclass[12pt,a4paper]{article}

\usepackage{a4}
\usepackage{amsfonts}
\usepackage{amssymb}
\usepackage{epsfig}
\usepackage{psfig}

\newcommand{\beqn}{\begin{eqnarray}}
\newcommand{\eeqn}{\end{eqnarray}}
\newcommand{\be}{\begin{equation}}
\newcommand{\ee}{\end{equation}}
\newcommand{\non}{\nonumber \\}

\begin{document} 

\title{
\begin{flushright}
\vspace{-3cm}
{\small HU-EP-00/61 \\
\vspace{-.3cm}
hep-th/0012156}
\end{flushright}
\vspace{3cm}
{\huge Type I Strings with ${F}$- and $B$-Flux}}
\author{} 
\date{}

\maketitle
\thispagestyle{empty}

\begin{center}
{\bf Ralph Blumenhagen}, \footnote{e-mail: blumenha@physik.hu-berlin.de} 
{\bf Boris K\"ors} \footnote{e-mail: koers@physik.hu-berlin.de}
and {\bf Dieter L\"ust} \footnote{e-mail: luest@physik.hu-berlin.de}
\\
\vspace{0.5cm}
Humboldt Universit\"at zu Berlin \\
{\small{Institut f\"ur Physik, Invalidenstr. 110, 10115 Berlin, Germany}} 
\vspace{2cm}
\end{center}

\begin{center}
{\bf Abstract} \\
\end{center}
We present non-supersymmetric toroidal compactifications of type I string
theory with both constant background NSNS two-form flux and 
non-trivial magnetic flux on the various D9-branes.   
The non-vanishing
$B$-flux admits four-dimensional mo\-dels with three generations of
chiral fermions in standard model like gauge groups.  
Additionally, we consider the orbifold $\mathbb{T}^4/\mathbb{Z}_2$, again with
both kinds of background flux present, leading to
non-supersymmetric as well as supersymmetric models 
in six dimensions.  
All models have T-dual descriptions as intersecting brane worlds. 

\clearpage

\section{Introduction}

In type I string theory there exist two known ways of achieving
chirality in the effective lower dimensional theory. On the one hand
one can compactify on curved spaces, in particular orbifolds, leading to
supersymmetric \cite{rang,rkakufour,rzwart, riban}
 and non-supersymmetric \cite{branesusy,aldaura}
models in six and four dimensions. 
On the other hand, as was first pointed out in \cite{bachas} 
and described in a pure 
stringy language in \cite{ours}, one can obtain  chiral spectra by introducing
D-branes with magnetic flux, or, in a T-dual interpretation,  
D-branes at angles \cite{angles1,angles2,pradisi,Asym,FHS}. \\

A nice geometric feature of such models is that the matter fields
are localized on the intersection locus of any two branes with their
multiplicity determined simp\-ly by the number of such intersection points.
In general, D-branes at angles not only lead to chirality but also break
supersymmetry completely at the string scale. In principle, this can be
reconciled by lowering the string scale to the electroweak regime in the way
of a large volume compactification.  
As was pointed out in \cite{AFIRU1,AFIRU2} for the case of type II theory, 
such intersecting brane world models have appealing phenomenological
implications, as for instance the absence of perturbative 
proton decay and a possible  hierarchy of the Yukawa couplings. 
However, for D-branes at angles there generically appear tachyons in the
spectrum, which in type II theory may trigger a decay into the vacuum. 
For this latter reason, it is worthwhile to consider type I 
models, where such a decay is forbidden due to the charge of the orientifold
planes. \\

Because of the relatively mild tadpole cancellation
conditions this ansatz leads to a plethora of solutions allowing
to construct semi-realistic models in a bottom-up approach.  
In addition to presenting the general framework, in \cite{ours} we pointed out
two subtleties for phenomenological applications of such models.
First, by introducing D-branes at angles into type I it
turned out to be impossible to get an odd number of matter generations due to
the symmetry of the D-brane spectrum under the world sheet parity.     
Second, with the internal space being a torus, $\mathbb{T}^D$, a
large extra dimension scenario was not compatible with chirality. 
This latter subtlety can be resolved by considering e.g. backgrounds of the form 
$\mathbb{T}^d\times \left( \mathbb{T}^{D-d}/\mathbb{Z}_3 \right)$ where the D-branes
wrap only $d/2$-cycles of the first torus and are pointlike on the
orbifold. This will work just as in the type II scenario of
\cite{AFIRU1,AFIRU2}, where, moreover, the first problem did not arise at
all. Another general feature of all the non-supersymmetric 
vacua is, that there remains an uncancelled NSNS tadpole which
leads to a shift in the background, as shown in \cite{dila}. \\ 

In this paper we continue to elaborate on type I strings with magnetic
flux and resolve the first subtlety by introducing discrete 
NSNS two-form flux \cite{rbfield}, as well. 
Via T-duality this corresponds
to a tilt of the dual torus. In section 2 we explain how the non-trivial gauge
background enters into the boundary states that provide the CFT description of
the D-branes with constant background flux on their world volume. In section 3
we present the generalized tadpole cancellation conditions and construct a
three generation left-right symmetric standard model. 
In the second part of the paper, section 4, we study  the 
$\mathbb{T}^4/\mathbb{Z}_2$ orbifold \cite{sagbi,rgimpol} of type I, again
with both kinds of fluxes present, and derive the
general form of the tadpole cancellation conditions, leading to a variety of
new six-dimensional models. They display a modified pattern of the
conventional reduction of the rank of the gauge group due to the $B$-field, 
revealing that the background $B$-field is not the reason for the reduction
of the rank of the D9-brane gauge group in the first place. Actually it is
better understood by noticing the multiple wrapping of the D9-branes on the
torus. Note, that partial results
for such models were already obtained in \cite{AADS}, where it was also
shown that there exist non-trivial configurations preserving
supersymmetry, which were related to (anti-)self dual background gauge fields. 

\section{D-branes on a $\mathbb{T}^2$ with background fluxes} 

In this section we discuss D-branes on a single two-dimensional torus
in the presence of background NSNS $B$- and magnetic ${F}$-fluxes. In particular, we
study how the data that describe the non-trivial gauge bundle on the torus,
the Chern number and Wilson lines, 
enter into the boundary states. This will be an essential ingredient to
compute 
the open string scattering diagrams needed to check consistency of
the type I models in such backgrounds.
Using this general approach we find a very intuitive mechanism
for the reduction of the rank of the gauge group in the case of
non-trivial $B$-flux. The actual cause 
is not the $B$-flux in the first place, but a multiple wrapping number
of the D-branes. \\

\begin{figure}[h] 
\begin{center}
\makebox[10cm]{
 \epsfxsize=14cm
 \epsfysize=6cm
 \epsfbox{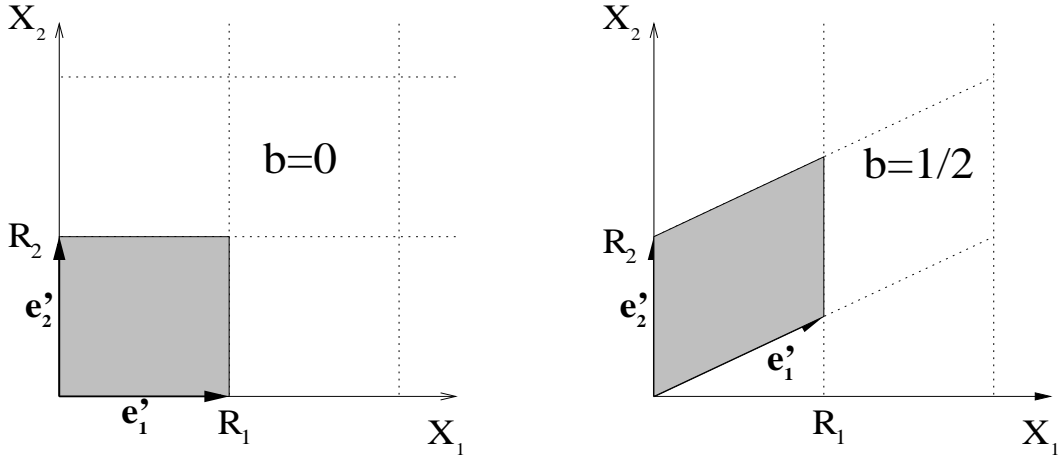}
}
\end{center}
\caption{Dual configurations}
\label{figlatticeB}
\end{figure}

Let us denote the coordinates by $x_{1,2}$ and the radii by $R_{1,2}$ and
characterize the torus $\mathbb{T}^2$ by its K\"ahler and complex structures 
\beqn
T = T_1 +iT_2 = b + i R_1 R_2,
\quad 
U = U_1 +iU_2 = i \frac{R_2}{R_1} .
\eeqn 
Since $U_1$ is a continuous modulus of the theory, for simplicity, we are
allowed to set it to zero. This choice enables us 
to remove the background NSNS field $B=b/T_2$ by a T-duality in the $x_2$ 
direction. The action of such a T-duality on the complex and K\"ahler
structure is
\beqn
T' =   -1/U ,\quad U' = -1/T.
\eeqn
It also transforms a D$p$-brane wrapping the entire torus into a D$(p-1)$-brane
stretching  along some 1-cycle. This is  evident from the mapping of the
boundary conditions 
\beqn
\partial_\sigma X_1 + {\cal F} \partial_\tau X_2 &=& 0 , \non
\partial_\sigma X_2 - {\cal F} \partial_\tau X_1 &=& 0 
\eeqn
of the D$p$-brane with magnetic flux ${\cal F}$ into 
\beqn
\partial_\sigma \left( X_1 + {\cal F} X_2 \right) &=& 0 , \non
\partial_\tau \left( X_2 - {\cal F} X_1 \right) &=& 0 ,
\eeqn
for a D$(p-1)$-brane at an angle $\phi ={\rm arctan}({\cal F})$ relative to
the $x_2$-axis. 
The T-dual tori for the cases $b=0$ and $b=1/2$ are displayed in
figure \ref{figlatticeB}. \\
%

\noindent
If we denote the 1-cycle wrapped by the D$(p-1)$-brane by 
\beqn
pe_1' + qe_2' \in H^1\left( \mathbb{T'}^{2}\right) ,
\eeqn
one can read off that
\beqn
{\cal F} = B+F = b\frac{R_2}{R_1 } +\frac{p}{q}\frac{R_2}{R_1 } .
\eeqn
The last equation relates the wrapping quantum numbers $p$ and $q$ to 
the T-dual
quantities that characterize the flux. In fact $p$ translates into
the number of times the D$p$-brane wraps the entire torus and $q$ to the first
Chern number of the gauge bundle, both taking values in $H^2\left
  ( \mathbb{T}^2\right)$.    
The condition that $p$ and $q$ are integers derives from the Dirac quantization
in the ``flux picture'' and from the requirement to have a well 
defined rational
wrapping of the D$(p-1)$-brane on the torus in the ``angle picture''. \\

An interesting point to notice is that for the case $b=1/2$ 
former D9-branes with $(p,q)=(2k,-k)$ 
now become D8-branes with even wrapping number on the cycle $e_1'$.  
Therefore, their minimal length is twice as long as compared to the case
$b=0$. Former D5-branes, which were point like on the torus now have general
integer wrapping on $e_2'$. Therefore, a single D9-brane in the 
presence of the $B$-flux carries twice the
charge of a D9-brane feeling no $B$-flux.
Contrarily, a D5-brane still carries one unit of RR-charge. 
This is the origin of the reduction of the rank of
the gauge group supported by the D9-branes: While the negative background
charge of the O9-plane stays unchanged, D9-branes carry a double amount of
charge, thus their number is halved. \\

In the following we will work in the ``angle picture'',
keeping in mind how it is related  to the ``flux picture''.
The open string Kaluza-Klein  and winding mass spectrum for 
coprime integers $p$ and $q$  is
\beqn \label{openKK}
M_{\rm open}^2(r,s) = 
\frac{r^2 + s^2\, (R_1 R_2)^2}{(q +b\, p)^2\,  (R_2)^2 + p^2\, (R_1)^2} , 
\eeqn
where the denominator is simply the squared length of the
brane along the wrapped cycle
\beqn \label{len}
L = \sqrt{(q +b\, p)^2\,  (R_2)^2 + p^2\, (R_1)^2} , 
\eeqn
while $R_1 R_2/L$ is the distance between two copies of the brane in the
elementary cell. Since loop and tree channel are related by a modular transformation,
the spectrum (\ref{openKK}) affects the normalization of the  
boundary state of a 
D$(p-1)$-brane with wrapping numbers $p$ and $q$
\beqn
\vert B\rangle _{(p,q)} = {1\over \sqrt{8}}{L
  \over \sqrt{R_1\, R_2}}
\Bigl(\vert NSNS, +\rangle - \vert NSNS, -\rangle
+  \vert RR, +\rangle + \vert RR, -\rangle \Bigr).
\eeqn
The non-trivial contribution of the bosonic modes along the
torus $\mathbb{T}^2$ is given by
\beqn
\vert B\rangle^{\rm bos}_{(p,q)} = \sum_{k,\omega} \exp 
\left( \sum_{n=1}^\infty{\frac{1}{n}\,
    e^{2i\phi}\, \alpha_{-n} \widetilde{\alpha}_{-n} +\ {\rm c.c.}\ } \right) 
    \vert k,\omega\rangle ,
\eeqn
with the zero-mode eigenstate $\vert k,\omega\rangle$ and a similar state for the
fermionic modes. The dependence of the boundary state on the Chern number of
the gauge bundle enters via $\phi$, the normalization factor
 and the zero modes $\vert k,\omega\rangle$.
Wilson lines can easily be implemented
by including phase factors into the zero-mode summations. Together the
boundary state encodes all the geometrical data. \\ 

Being equipped with the boundary states of wrapped branes, 
it is a straightforward exercise to compute the various one-loop
amplitudes. In particular, transforming the tree channel annulus amplitude 
for two non-parallel D-branes into the loop channel reveals an extra 
multiplicative factor in front of the amplitude, 
\beqn \label{intnr}
I_{\mu \nu} = p_\mu q_\nu - q_\mu p_\nu ,
\eeqn 
which is nothing else than the intersection number of the two D-branes
on the torus $\mathbb{T}^2$.
Thus, as we are used to in string theory, a pure conformal field theory
computation allows us to compute topological data. 
Taking the  intersection numbers (\ref{intnr}) into account is  essential 
when one computes 
the massless open string spectrum. Similarly, when considering models
with an additional orbifold action, the twisted sector components
of the  boundary states contain all information about
the number of intersection points invariant under the
orbifold action. \\

\section{Non-supersymmetric toroidal models with $B$- and $F$-flux} 

In this section we generalize the previous work \cite{ours}
on type I strings on a torus with magnetic background ${F}$-flux to the
most general combination of ${F}$- and NSNS $B$-flux. 
To begin with, let us  fix some  notation.
The four- or six-dimensional torus is assumed  to factorize according to 
\beqn
\mathbb{T}^{2d} = {\mbox{\huge $\otimes$}}_{j=1}^d{\mathbb{T}_{(j)}^2} 
\eeqn
with vanishing $U_1^{(j)}$ for all the two-dimensional tori. 
As usual, due to the $\Omega$ projection the NSNS $B^{(j)}$-field is  
constrained to take the discrete values 
$b^{(j)}=0,1/2$. 
Now, we add $K$ stacks   of ${\rm N}_\mu$ D9$_\mu$-branes  
with magnetic $F_\mu$-flux turned on and look for 
configurations cancelling the RR tadpole of the Klein-bottle. \\

As explained above, this configuration is T-dual to a configuration of
D$(9-d)$-branes intersecting at angles on a torus without background 
fluxes but with non-trivial complex structure and relative angles between the
branes. Thus, in the T-dual model we have a purely
geometric picture of what is going on.
Note, that the world-sheet parity $\Omega$ gets transformed
into $\Omega {\cal R}$ by this T-duality, where  ${\cal R}$ denotes
a reflection of all
the $x_2^{(j)}$ directions of the $T_{(j)}^2$. Thus, for $b=0$, 
$\Omega$ maps a
$(p,q)$ brane to $(p,-q)$ brane, whereas for $b=1/2$ it sends a $(p,q)$ brane
to a $(p,-p-q)$ brane. Overall, we need to have $2K$ stacks of branes counting
branes and their mirror branes separately. \\

\subsection{Tadpole cancellation}

We will not present a comprehensive computation of the one-loop tadpole
contributions but merely state the differences as compared to the earlier
work with $b=0$ in \cite{ours}. Indeed, the explicit expressions for the
amplitudes on the torus can be obtained from those on 
$\mathbb{T}^4/\mathbb{Z}_2$, as given in
the next chapter, by simply omitting the contributions arising  from 
the orbifold
insertion in the loop channel trace. \\

The presence of $b^{(j)}=1/2$ concretely
enters the amplitudes at three distinguished points. First, the Kaluza-Klein
 and
winding mass spectrum in (\ref{openKK}), second the number of 
intersection points of two branes and finally the number of
such intersections which are invariant under $\Omega$.
In all three cases the modification can be
summarized by noting that the winding numbers $(n^{(j)}_\mu,m_\mu^{(j)})$ 
for the $b^{(j)}=0$ case are replaced by
$(p_\mu^{(j)},q_\mu^{(j)}+p_\mu^{(j)}/2)$.  
The origin of this is exclusively found in the different normalization of the
boundary state due to the zero-mode spectrum (\ref{openKK}). 
Since the  Klein-bottle amplitude remains unchanged, the tadpole
cancellation conditions in six dimensions  read
\beqn
& &\sum_{\mu=1}^K {{\rm N}_\mu \prod_{j=1}^2{p_\mu^{(j)}}} = 16, \non  
& & \sum_{\mu=1}^K {{\rm N}_\mu \prod_{j=1}^2{\left(q_\mu^{(j)}+
      b^{(j)}\, p_\mu^{(j)}\right) }} = 0 ,
\eeqn
while those in four dimensions are
\beqn
& & \sum_{\mu=1}^K {{\rm N}_\mu\,  \prod_{j=1}^3{p_\mu^{(j)}}} = 16, \non 
& & \sum_{\mu=1}^K {{\rm N}_\mu\,  p_\mu^{(1)} \prod_{j=2,3}
{\left(q_\mu^{(j)}+
       b^{(j)}\, p_\mu^{(j)}\right) }} = 0 , \non 
& & \sum_{\mu=1}^K {{\rm N}_\mu\,  p_\mu^{(2)} 
 \prod_{j=1,3}{\left(q_\mu^{(j)}+
        b^{(j)}\, p_\mu^{(j)}\right) }} =0 , \non
& & \sum_{\mu=1}^K {{\rm N}_\mu\, p_\mu^{(3)} \prod_{j=1,2}{\left(q_\mu^{(j)}+
       b^{(j)}\, p_\mu^{(j)}\right) }} =0 .
\eeqn
Remember that pure D9-branes without any flux correspond to 
D$(9-d)$-branes with
even $p$, so that a theory with only D9-branes has a gauge group of rank
$16/{\rm rk}(B)$. Apparently, this rank reduction is not a direct 
consequence of the background $B$-field, but only follows from the doubled
wrapping number. The general spectrum of massless chiral fermions is formally
the same as given in \cite{ours}, with multiplicities derived from
$(p_\mu^{(j)},q_\mu^{(j)}+p_\mu^{(j)}/2)$ instead of
$(n^{(j)}_\mu,m_\mu^{(j)})$. \\  

\subsection{A three generation model}

In \cite{ours}  we studied in a bottom-up approach toroidal models with 
vanishing $B$-field and found that the net generation number of chiral
fermions was forced to be even.  Thus, it was impossible to construct a
standard model like spectrum with three generations. \\

In this section we will show that this constraint is weakened by turning
on background $B$-form flux respectively  tilting the torus in the T-dual 
picture. In particular, we present a four-dimensional left-right symmetric
standard model with three generations. 
To this end, we choose four stacks of D6-branes with numbers $N_1=3$,
$N_2=N_3=2$ and $N_4=1$. Moreover, only on the second torus 
we turn on $b^{(2)}=1/2$. 
The following choice of wrapping numbers
\beqn 
p^{(j)}_\mu=\left(\matrix{ 1 & 1 & 1 & 1 \cr
                           1 & 1 & 1 & 1 \cr
                           3 &  1 & 1 & 3 \cr} \right),
\quad\quad
             q^{(j)}_\mu=\left(\matrix{ 0 & 1 & 1 & 0 \cr
                                        0 & 1 & -2 & -2 \cr
                                       1 &  0 & 0 & 1 \cr} \right) 
\eeqn
satisfies the tadpole cancellation conditions and leads to the gauge group
$U(3)\times U(2)\times U(2)\times U(1)$ and 
the following chiral massless spectrum \\

\begin{table}[h]
\caption{Chiral massless spectrum}
\begin{center} 
\begin{tabular}{|l|l|}
\hline 
$SU(3)\times SU(2)_L\times SU(2)_R\times U(1)^4$  & number\\
\hline 
$({\bf 3},{\bf 2},{\bf 1})_{(1,1,0,0)}$ &   $2$ \\
$({\bf 3},{\bf 2},{\bf 1})_{(1,-1,0,0)}$ &  $1$ \\
\hline
$(\overline{\bf 3},{\bf 1},{\bf 2})_{(-1,0,1,0)}$ &  $2$ \\
$(\overline{\bf 3},{\bf 1},{\bf 2})_{(-1,0,-1,0)}$ &  $1$ \\
\hline
$({\bf 1},{\bf 2},{\bf 1})_{(0,-1,0,1)}$ &  $3$ \\
\hline
$({\bf 1},{\bf 1},{\bf 2})_{(0,0,-1,-1)}$ &  $3$ \\
\hline
\end{tabular}
\end{center}
\end{table}

\noindent
Computing the mixed $G^2\times U(1)$ anomalies, one realizes that
only two of the four $U(1)$ factors are anomaly-free. The remaining
two should get a mass by a generalized Green-Schwarz mechanism involving
NSNS scalars \cite{AFIRU1}.
In particular
\beqn
  U(1)_{B-L}={1\over 3}\left( U(1)_1-3\, U(1)_4\right) 
\eeqn
is one of the anomaly-free abelian gauge groups. The model is designed such
that the spectrum comes with the correct quantum numbers to use it as the
$U(1)$ for $B-L$. The final chiral massless spectrum is shown in table 2.

\begin{table}[h]
\caption{Chiral massless spectrum}
\begin{center} 
\begin{tabular}{|l|l|}
\hline 
$SU(3)\times SU(2)_L\times SU(2)_R\times U(1)^2$  & number\\
\hline 
$({\bf 3},{\bf 2},{\bf 1})_{(1/3,1)}$ &   $2$ \\
$({\bf 3},{\bf 2},{\bf 1})_{(1/3,-1)}$ &  $1$ \\
\hline
$(\overline{\bf 3},{\bf 1},{\bf 2})_{(-1/3,1)}$ &  $2$ \\
$(\overline{\bf 3},{\bf 1},{\bf 2})_{(-1/3,-1)}$ &  $1$ \\
\hline
$({\bf 1},{\bf 2},{\bf 1})_{(-1,-1)}$ &  $3$ \\
\hline
$({\bf 1},{\bf 1},{\bf 2})_{(1,1)}$ &  $3$ \\
\hline
\end{tabular}
\end{center}
\end{table}

\noindent
Depending on the actual values of the angles, we get tachyons in
bi-fundamental representations of the gauge group, which
trigger a decay of in each case two D-branes at angles into
a supersymmetric 3-cycle \cite{witten,bbh}.
This model should serve as an example to show how easily standard model like
spectra can be constructed. 

\section{Type I strings on K3 with background fluxes} 

We now consider type I string theory  compactified to six dimensions on 
K3 at its $\mathbb{T}^4/\mathbb{Z}_2$ orbifold point.   
As in the previous section we allow  the presence of constant NSNS 
$B$-flux in
the bulk as well as constant magnetic ${F}$-flux on the D9-branes. 
Whereas partial results have already been presented in \cite{AADS},
our aim is to treat the problem quite generally again employing the ``angle
picture''. The torus is defined as in the previous sections and 
$\mathbb{Z}_2=\{ 1,\Theta\}$. 

\subsection{Tadpole cancellation}

In the following we compute explicitly 
the contributions to the one-loop tadpoles. In addition to the different
normalizations of the boundary states describing the D-branes, we now also have
to face a different normalization of the O5-plane cross-cap state by a factor
of $2^{-{\rm rk}(B)/2}$, which affects the Klein Bottle and the
M\"obius strip. In the
following we present the results for the tree channel amplitudes and extract
the contributions to the massless tadpoles.  
The Klein bottle in  tree channel reads 
\beqn 
\tilde{\cal K} &=& \int_0^\infty  dl\ \left(  
    \langle \Omega {\cal R}\vert e^{-\pi l {\cal H}_{\rm cl}}
    \vert \Omega {\cal R} \rangle + \langle \Omega {\cal R}\Theta\vert e^{-\pi l {\cal H}_{\rm cl}}
    \vert \Omega {\cal R}\Theta \rangle \right) \non
&=& 2^6\, c\, (1-1) 
\int_0^\infty dl\ \left[ 
\frac{\vartheta \left[ 1/2 \atop 0 \right]^4}{\eta^{12}} 
\left( \prod_{j=1}^2 { R^{(j)}_1\over R^{(j)}_2}
       \prod_{i,j=1}^2   \sum_{r\in \mathbb{Z}} \exp\left(-4\pi l r^2
        \left( R^{(j)}_i \right)^2 \right)  \right. \right. \non
& & \hspace{3cm} + \left. \left. 
 \prod_{j=1}^2 {1\over 16^{b^{(j)}}} { R^{(j)}_2\over R^{(j)}_1} 
  \prod_{i,j=1}^2  
    \sum_{s\in \mathbb{Z}} \exp\left( 
 {\frac{-4\pi l s^2}{16^{ b^{(j)}} 
\left(R^{(j)}_i\right)^2 }} \right)
\right) \right] , 
\eeqn 
leading to the following contribution to the massless RR tadpole
\beqn
\tilde{\cal K} \sim 2^{10} \left( \prod_{j=1}^2 { R^{(j)}_1\over R^{(j)}_2}
    + \prod_{j=1}^2 {1\over 16^{b^{(j)}}} { R^{(j)}_2\over R^{(j)}_1}
   \right) \int_0^\infty{dl} .
\eeqn
As is well known, 
the charge of the former orientifold O9-plane state, corresponding to 
$\vert \Omega {\cal R}\rangle$, remains unchanged, while the charge of the
former O5-planes $\vert \Omega {\cal R} \Theta \rangle$ is reduced by the $B$-flux. \\

Next we have to compute the annulus amplitude, which receives contributions
from all the open strings stretching among the various
D7-branes. 
We denote by $\gamma_{\Theta,\mu}$ the action of the orbifold generator 
on the Chan-Paton indices of a string
ending on a D9$_\mu$-brane. We let
$\Delta_{\mu}^{(i)}=0,1$ count if the D9$_\mu$-brane passes through the 
$i$th $\mathbb{Z}_2$ fixed point, $i=1,...,16$. Note, that
each D7-brane runs exactly through four different fixed points. 
The RR tree channel annulus amplitude of strings with both ends on a
single D7-brane in  loop channel reads 
\beqn
\tilde{\cal A}_{\mu \mu} &=& \int_0^\infty dl\ \langle D_\mu,RR \vert 
e^{-\pi l {\cal H}_{\rm cl}} \vert D_\mu,RR \rangle \non
&=& 2^{-4} c
\int_0^\infty dl\ \left[ 
\frac{\vartheta \left[ 1/2 \atop 0 \right]^4}{\eta^{12}} \,
{\rm N}_\mu^2\, \prod_{j=1}^2 \left( 
  \frac{\left(L^{(j)}_\mu\right)^2}{R_1^{(j)}R_2^{(j)}}
\sum_{r,s\in \mathbb{Z}^2}{e^{-\pi l \tilde{M}^2_{\mu,j}
         }} 
\right) +
 \right.  \non
& & \hspace{4cm} \left. 
 4 \frac{\vartheta \left[ 1/2 \atop 0 \right]^2\vartheta \left[ 0 \atop 0
  \right]^2}
{\eta^6\vartheta \left[ 0 \atop 1/2 \right]^2}
\sum_{i=1}^{16}{\Delta_{\mu}^{(i)} 
               {\rm tr}\left( \gamma_{\Theta,\mu} \right)^2 } 
\right] .
\eeqn
from which one can easily read off the massless untwisted and twisted 
tadpoles. Note that the boundary states now also include twisted components
which insure the orbifold projection in the loop channel. The annulus
amplitude between two different branes in  tree channel is
\beqn
\tilde{\cal A}_{\mu \nu} &=& \int_0^\infty dl\ \left( \langle D_\mu,RR 
\vert e^{-\pi l {\cal H}_{\rm cl}}
    \vert D_\nu,RR \rangle +  \langle  D_\nu,RR 
\vert e^{-\pi l {\cal H}_{\rm cl}} \vert D_\mu,RR \rangle \right) \\
&=& 2^{-1} c\,  
\int_0^\infty dl\ \left( {\rm N}_\mu {\rm N}_\nu I_{\mu \nu} 
     \frac{\vartheta \left[ 1/2 \atop 0 \right]^2 \prod_{j=1}^2{\vartheta
         \left[ 1/2 \atop (\phi_\mu^{(j)}-\phi_\nu^{(j)})/\pi \right]^2}}
{\eta^6 \prod_{j=1}^2{\vartheta \left[ 1/2 \atop 1/2 +
      (\phi_\mu^{(j)}-\phi_\nu^{(j)})/\pi \right]^2}}   
\right.  \non 
& & \hspace{1.5cm} + \left.  
\frac{\vartheta \left[ 1/2 \atop 0 \right]^2 \prod_{j=1}^2{\vartheta
         \left[ 0 \atop (\phi_\mu^{(j)}-\phi_\nu^{(j)})/\pi \right]}}
{\eta^6 \prod_{j=1}^2{\vartheta \left[ 0 \atop 1/2 +
      (\phi_\mu^{(j)}-\phi_\nu^{(j)})/\pi \right]}}   
\sum_{i=1}^{16}{ \Delta_{\mu}^{(i)} \Delta_{\nu}^{(i)} 
    {\rm tr}\left( \gamma_{\Theta,\mu} \right) {\rm tr}\left
    ( \gamma_{\Theta,\nu} \right) } 
\right) \nonumber  
\eeqn
where the number of common fixed points can be written as
\beqn
\sum_{i=1}^{16}  \Delta_{\mu}^{(i)} \Delta_{\nu}^{(i)} =
  \prod_{j=1}^2 \left[ 1+{1\over 4}\left(\sum_{\delta=0}^1 
  e^{\pi i \delta (p^{(j)}_\mu -p^{(j)}_\nu)} \right)
  \left(\sum_{\epsilon=0}^1 
  e^{\pi i \epsilon (q^{(j)}_\mu -q^{(j)}_\nu)} \right) \right].
\eeqn
The M\"obius amplitude receives  only contributions from strings 
stretching between branes and their images under $\Omega {\cal R}$. 
In tree channel one gets
\beqn
\widetilde{\cal M}_\mu &=&  \int_0^\infty dl\ \left( \langle D_\mu+D_{\mu'},RR 
\vert e^{-\pi l {\cal H}_{\rm cl}}
    \vert \Omega {\cal R} +\Omega {\cal R}\Theta,RR\rangle + \right. \non
  &\phantom{=}& \quad\quad\quad \left. \langle  
   \Omega {\cal R} +\Omega {\cal R}\Theta,RR  \vert e^{-\pi l {\cal H}_{\rm cl}}
    \vert  D_\mu+D_{\mu'},RR  \rangle  \right) \non
&=& \pm 2^5 c\, \int_0^\infty dl\  {\rm N}_\mu\, 
\frac{\vartheta \left[ 1/2 \atop 0 \right]^2}{\eta^6} 
\left(\prod_{j=1}^2 \left( q^{(j)}_\mu + b^{(j)} p^{(j)}_\mu \right)
\frac{\prod_{j=1}^2{\vartheta \left[ 1/2 \atop -\phi_\mu^{(j)}/\pi \right]}}
    {\prod_{j=1}^2{\vartheta \left[ 1/2 \atop 1/2-\phi_\mu^{(j)}/\pi \right]}}
\right.   \non
& &  \hspace{4cm} + \left.\prod_{j=1}^2 p^{(j)}_\mu 4^{-b^{(j)}}\    
\frac{\prod_{j=1}^2 {\vartheta 
\left[ 1/2 \atop 1/2-\phi_\mu^{(j)}/\pi \right]}}
{\prod_{j=1}^2{\vartheta \left[ 1/2 \atop -\phi_\mu^{(j)}/\pi \right]}}
\right) . 
\eeqn
which leads to the following contribution to the
massless RR tadpole
\beqn
\tilde{\cal M}_\mu  = \pm 2^{7}\, {\rm N}_\mu \left( 
  \prod_{j=1}^2 p_\mu^{(j)}{R_1^{(j)}\over R_2^{(j)}} + 
   \prod_{j=1}^2 \left( q^{(j)}_\mu + b^{(j)} p^{(j)}_\mu \right)
 {R_2^{(j)}\over R_1^{(j)}}\, 4^{-b^{(j)}}  \right) \int_0^\infty{dl} .
\eeqn
Adding up all the different contributions for all possible open
strings  one gets the general tadpole cancellation conditions 
\beqn
& & \sum_{\mu=1}^K  {\rm N}_\mu \prod_{j=1}^2 p_\mu^{(j)} = 16, \non  
& & \sum_{\mu=1}^K  {\rm N}_\mu \prod_{j=1}^2 
\left( q^{(j)}_\mu + b^{(j)} p^{(j)}_\mu \right) = 16 \prod_{j=1}^2 
   4^{-b^{(j)}}, \non 
& & \sum_{\mu=1}^{2K} {\Delta_\mu^{(i)} \,
        {\rm tr} \left( \gamma_{\Theta,\mu} \right)} =0, \quad i=1,...,16.  
\eeqn
where in the first two equations the sum runs only  
over the  $K$ D7-branes without counting their mirror
branes separately. In the third equation the sum runs over all $2K$ branes. \\
 
Note, that the presence of the $B$-flux does not necessarily imply a 
reduction of
the rank of the gauge group. As we will show in the next subsection, 
we can even have ${\rm rk}(B)=4$ together with a
gauge group of rank 16. 

\subsection{Massless spectra}

In this subsection we will derive the generic form of the
massless spectrum distinguishing among the different intersection
points. 
The generic solution to the twisted tadpole cancellation is   simply
\beqn
{\rm tr}\left( \gamma_{\Theta,\mu} \right) =0 , \quad {\rm for\ all}\ \mu.  
\eeqn
This implies a gauge group 
\beqn
\prod_{\mu=1}^K  U(N_\mu/2)\times U(N_\mu/2).
\eeqn
For convenience we define $M_\mu =N_\mu/2$. 
The action of $\Omega{\cal R}$ on Chan-Paton indices is 
\beqn
\gamma_{\Omega{\cal R},\mu}
\left( \begin{array}{cc} 
A_1 & A_2 \\ A_3 & A_4 \end{array} \right)^T
\gamma_{\Omega{\cal R},\mu}^{-1}= 
\left( \begin{array}{cc} 
A_4^T & A_2^T \\ A_3^T & A_1^T \end{array} \right) ,
\eeqn
the $A_i$ being $N_\mu\times N_\mu$ matrices, $A_1$ for $\mu\mu$ strings,
$A_2$ for $\mu\mu'$ strings, etc. The reflection $\Theta$ leaves
all individual branes invariant and thus acts separately on the $N_\mu\times
N_\mu$ factors
\beqn
\gamma_{\Theta,\mu} \left( \begin{array}{cc} 
B_1 & B_2 \\ B_3 & B_4 \end{array} \right) 
\gamma^{-1}_{\Theta,\mu} = \left( \begin{array}{cc} 
B_1 & -B_2 \\ -B_3 & B_4 \end{array} \right) .
\eeqn
Chiral massless states are localized 
at the intersection locus of any two D7-branes and the representation they are
transforming in under the gauge group depends crucially on the behaviour
of the intersection point under $\Omega{\cal R}$ and $\Theta$. 
Moreover, a negative
intersection number flips the chirality of a massless fermion. 
Taking all this into account the the matter content of the various open 
string sectors is summarized in table 3. \\

\clearpage

\begin{table}[h]
\caption{Chiral massless spectra}
\label{tablespectrum}
\begin{center} 
\begin{tabular}{|c|c|c|c|}
\hline
Sector & Spin & $(\Omega{\cal R}(\Theta), \Theta)$ & Matter \\ 
\hline
$\mu\mu$ & $(2,1)$ & $(-,\Theta)$ & 2$(({\bf Adj, 1})+ ({\bf 1,Adj})) $ \\
$\mu\mu$ & $(1,2)$ & $(-,\Theta)$ & 2$({\bf M_\mu, M_\mu})+conj.$ \\
\hline
$\mu\mu'$ & $(1,2)$ & $(\Omega{\cal R}(\Theta),\Theta)$ & $({\bf A_\mu ,1}) + ({\bf 1,
  A_\mu})+conj. $ \\
$\mu\mu'$ & $(2,1)$ & $(\Omega{\cal R}(\Theta),\Theta)$ & $({\bf M_\mu ,M_\mu}) +conj. $ \\
$\mu\mu'$ & $(2,1),(1,2)$ & $(\Omega{\cal R}(\Theta),-)$ & $({\bf A_\mu ,1}) + ({\bf 1,
  A_\mu})+({\bf M_\mu ,M_\mu})+conj. $ \\
$\mu\mu'$ & $(2,1),(1,2)$ & $(-,-)$ & $({\bf A_\mu ,1}) + ({\bf 1,
  A_\mu})+({\bf S_\mu ,1}) + ({\bf 1,
  S_\mu})+ $ \\
& & & $2({\bf M_\mu ,M_\mu})+conj. $ \\
\hline
$\mu\nu,\mu\nu'$ & $(1,2)$ & $(-,\Theta)$  & $ ({\bf M_\mu,1,M_\nu,1})+
({\bf 1,M_\mu,1,M_\nu}) +conj.$ \\
$\mu\nu,\mu\nu'$ & $(2,1)$ & $(-,\Theta)$  & $ ({\bf M_\mu,1,1,M_\nu})+
({\bf 1,M_\mu,M_\nu,1}) +conj.$ \\
$\mu\nu,\mu\nu'$ & $(1,2),(2,1)$ & $(-,-)$  & $ ({\bf M_\mu,1,M_\nu,1})+
({\bf 1,M_\mu,1,M_\nu})+$ \\
&  & & $({\bf M_\mu,1,1,M_\nu})+({\bf 1,M_\mu,M_\nu,1}) +conj.$ \\
\hline
\end{tabular}
\end{center}
\end{table}

\noindent
We always count single fermions, which, whenever the configurations become
supersymmetric, may combine into proper supermultiplets. 
The column  $(\Omega{\cal R}(\Theta), \Theta)$ denotes whether the 
intersection point is invariant under $\Omega{\cal R}$ or $\Omega{\cal R}
\Theta$ and $\Theta$, respectively, which then requires to include the proper
projections. \\

When any two branes $\mu$ and $\nu$ pass through the same set of fixed points
the conditions for the twisted tadpoles can be expressed in terms of 
${\rm tr}\left( \gamma_{\Theta,\mu} + \gamma_{\Theta,\nu} \right)$. 
In the case of
$\nu=\mu'$ this results in a  larger gauge group $U(N_\mu)$, which happened
accidently in all the examples studied in \cite{AADS}, where only the special
cases $(p,q)\in (2,\mathbb{Z})$ of non-trivial $F$-fluxes in combination with
$B$-fields have been discussed. 

\subsection{Examples}

\begin{figure}[h] 
\begin{center}
\makebox[10cm]{
 \epsfxsize=14cm
 \epsfysize=21cm
 \epsfbox{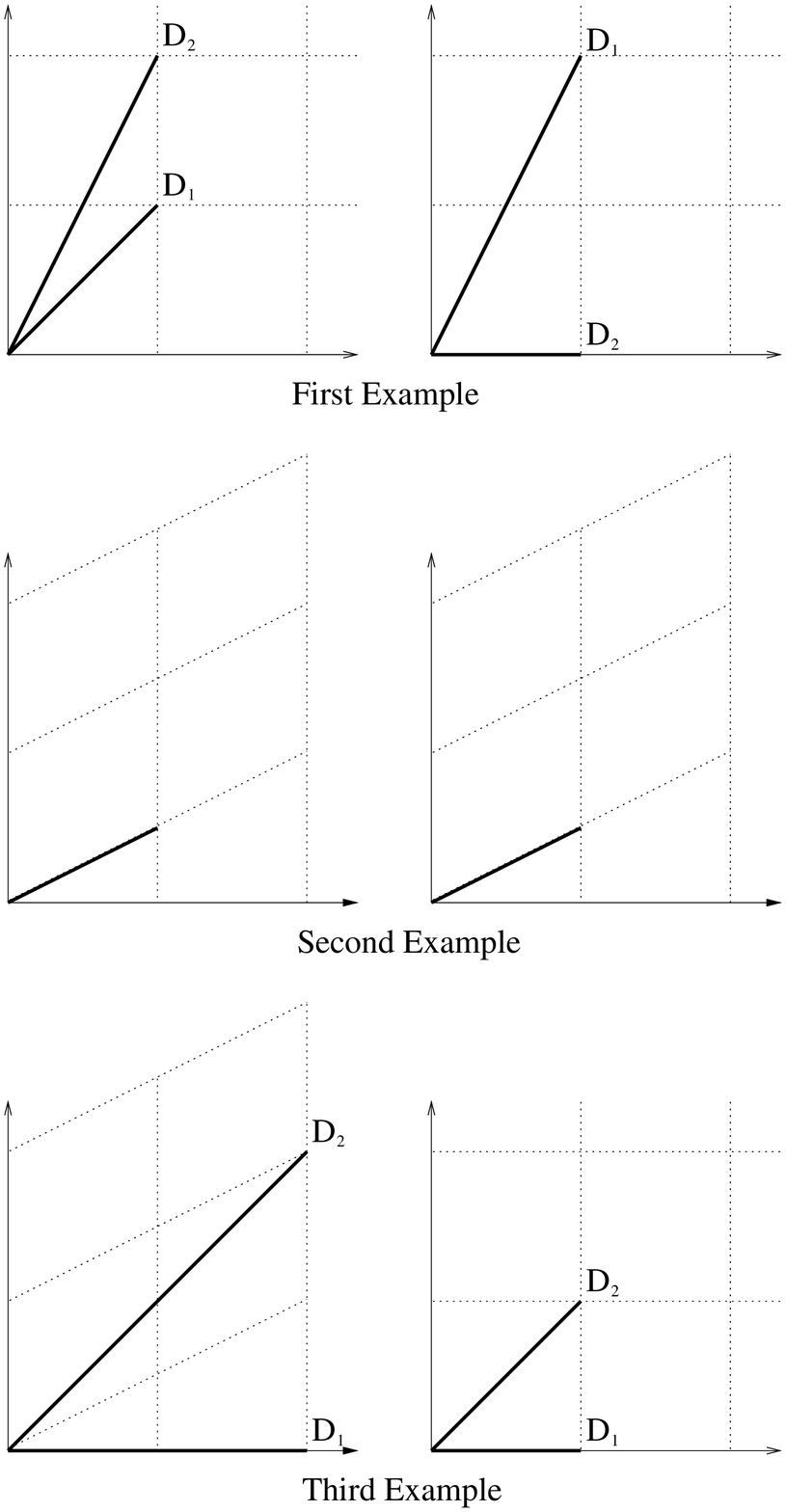}
}
\end{center}
\caption{D7-brane configurations of 6D examples}
\label{figexamples}
\end{figure}

In this subsection we will discuss a few examples in some more detail
and, as expected, we will find that all these models are anomaly free
in six dimensions. The configurations of D7-branes are displayed in figure
\ref{figexamples}.  

\subsubsection{Example 1}

We choose two stacks of branes ($K=2$) with $N_1=N_2=8$ and
vanishing $B$-field. 
The following choice of wrapping numbers
\beqn 
p^{(j)}_\mu=\left(\matrix{ 1 & 1  \cr
                           1 & 1 \cr} \right),
\quad\quad
             q^{(j)}_\mu=\left(\matrix{ 1 & 2  \cr
                                        2 & 0 \cr} \right) 
\eeqn
leads to the chiral massless spectrum shown in table 4. \\

\begin{table}[h]
\caption{Spectrum of example 1}
\begin{center} 
\begin{tabular}{|c|c|c|}
\hline 
sector & spin & $U(4)\times U(4) \times U(4)\times U(4)$  \\
\hline 
11,22  & (2,1) & $2({\bf Adj},{\bf 1},{\bf 1},{\bf 1})+cycl. $     \\
11    & (1,2) & $2({\bf 4},{\bf 4},{\bf 1},{\bf 1})+conj. $    \\
22    & (1,2) & $2({\bf 1},{\bf 1},{\bf 4},{\bf 4})+conj. $    \\
\hline
11' & (1,2) & $6({\bf A},{\bf 1},{\bf 1},{\bf 1})+
               6({\bf 1},{\bf A},{\bf 1},{\bf 1})+
               2({\bf 4},{\bf 4},{\bf 1},{\bf 1}) +conj.$    \\
22' & (1,2) & $3({\bf 1},{\bf 1},{\bf A},{\bf 1})+
               3({\bf 1},{\bf 1},{\bf 1},{\bf A})+
               1({\bf 1},{\bf 1},{\bf 4},{\bf 4}) +conj.$    \\
   & (2,1) & $3({\bf 1},{\bf 1},{\bf 4},{\bf 4})+
               1({\bf 1},{\bf 1},{\bf A},{\bf 1})+
               1({\bf 1},{\bf 1},{\bf 1},{\bf A}) +conj.$    \\
\hline
12 & (2,1) & $2({\bf 4},{\bf 1},{\bf 1},{\bf 4})+
               2({\bf 1},{\bf 4},{\bf 4},{\bf 1})+conj. $\\
12' & (1,2) & $4({\bf 4},{\bf 1},{\bf 4},{\bf 1})+
               4({\bf 1},{\bf 4},{\bf 1},{\bf 4})+
               2({\bf 4},{\bf 1},{\bf 1},{\bf 4})+
               2({\bf 1},{\bf 4},{\bf 4},{\bf 1})+conj. $\\
\hline
\end{tabular}
\end{center}
\end{table}

\noindent
Together with the 20 hypermultiplets and 1 tensormultiplet from the
closed string sector the spectrum in table 4 indeed satisfies
$F^4$ and $R^4$ anomaly cancellation. In contrast to the toroidal
case, the formal vanishing of the intersection number $I_{22'}$ does
not imply a non-chiral spectrum in this open string sector due to the orbifold \\
projection. 
%

\subsubsection{Example 2}

We choose only one  stack of D7-branes ($K=1$) with $N_1=N_2=16$ and 
wrapping numbers $(p_1,q_1)=(p_2,q_2)=(1,0)$ on the two tori.
With a  $B$-field of rank four we obtain the simple anomaly free massless
spectrum in table 5. \\

\begin{table}[h]
\caption{Spectrum of example 2}
\begin{center} 
\begin{tabular}{|c|c|c|}
\hline 
sector & spin & $U(8)\times U(8) $  \\
\hline 
11  & (2,1) & $2({\bf Adj},{\bf 1})+2({\bf 1},{\bf Adj})$     \\
11    & (1,2) & $2({\bf 8},{\bf 8}) +conj.$    \\
\hline
11' & (1,2) & $({\bf A},{\bf 1})+
               ({\bf 1},{\bf A})  +conj.$    \\
\hline
\end{tabular}
\end{center}
\end{table}

\noindent
This is an example of a model where the rank of the gauge group is not reduced
by rk$(B)=4$ but only by a factor of 2. 
Together with the 7 tensormultiplets and 14 hypermultiplets from
the closed string sector the model satisfies anomaly cancellation. 
Note, that for appropriate choice of the radii of the two tori 
the flux is self-dual and the model becomes supersymmetric. 

\subsubsection{Example 3}

Finally, we discuss a model which was also considered in \cite{AADS}.
We choose two stacks of branes ($K=2$) with $N_1=N_2=4$ and
non-zero  $B$-field on the first torus $b^{(1)}=1/2$ only. 
A solution to the tadpole cancellation conditions
is given by the wrapping numbers
\beqn 
p^{(j)}_\mu=\left(\matrix{ 2 & 1  \cr
                           2 & 1 \cr} \right),
\quad\quad
             q^{(j)}_\mu=\left(\matrix{ -1 & 0  \cr
                                        1 & 1 \cr} \right), 
\eeqn
where the first D7-brane is stretched along the $x_1$ axis of the
two tori, i.e. corresponds to a D9-brane without flux. 
Therefore, the mirror brane is of the same kind as the
orginal brane and effectively we have a $U(4)$ gauge group, instead of $U(2)$. 
Moreover, the second  brane and its mirror brane
run through the same set of fixed points. Thus, we can
satisfy the twisted tadpole condition by choosing
\beqn
     {\rm tr} \left( \gamma_{\Theta,2} \right) =-
   {\rm tr} \left( \gamma_{\Theta,2'} \right).
\eeqn
Therefore, there is no orbifold projection on the Chan-Paton labels for the
gauge group living on the D$_2$-branes. 
Computing all the intersection numbers and taking the transformation
properties of the intersection points into account we obtain the
massless spectrum displayed in table 6. \\

\begin{table}[h]
\caption{Spectrum of example 3}
\begin{center} 
\begin{tabular}{|c|c|c|}
\hline 
sector & spin & $U(4)\times U(4)$  \\
\hline 
11,22  & (2,1) & $2({\bf Adj},{\bf 1})+2({\bf 1},{\bf Adj})$     \\
11    & (1,2) & $2({\bf A},{\bf 1})+conj. $   \\
\hline
22' & (1,2) & $8({\bf 1},{\bf A})+
               2({\bf 1},{\bf S})+conj. $   \\
\hline
12 & (1,2) & $4({\bf 4},{\bf 4})+ conj. $ \\
\hline
\end{tabular}
\end{center}
\end{table}

\noindent
This completely agrees with the result obtained in \cite{AADS}.

\section{Conclusions}

In this paper we have studied type I models with D9-branes
with background $B$- and $F$-flux, respectively T-dual models 
with D7- or D6-branes at angles.
In contrast to the case with vanishing $B$-flux, here we encountered  no
obstruction to obtain semi-realistic three generation models in a bottom-up approach. 
Moreover, we have studied the  $\mathbb{T}^4/\mathbb{Z}_2$ orbifold model with
fluxes in generality and obtained a whole plethora of anomaly-free 
non-supersymmetric as well as supersymmetric models in six space-time
dimensions. 

\clearpage

\bibliography{articles}
\bibliographystyle{unsrt}

\end{document}